\documentclass{aa} 
\usepackage{graphicx}
\begin{document}

\title{Relativistic Parker winds with variable effective polytropic index}

  \titlerunning{}

  \author{Z. Meliani
           \inst{1,2}
   \and  C. Sauty
           \inst{1,2}
   \and  K. Tsinganos
           \inst{3}
   \and  N. Vlahakis
           \inst{3}
          }

   \offprints{Z. Meliani\\ (zakaria.meliani@obspm.fr)}

   \institute
         {Observatoire de Paris, L.U.Th., F-92190 Meudon, France
    \and Universit\'e de Paris 7, APC, 2 place Jussieu, F-75005 Paris, France
    \and IASA and Section of Astrophysics, Astronomy \& Mechanics,
         Department of Physics, University of Athens,
         Panepistimiopolis GR-157 84, Zografos, Greece
}
   \date{Received ... / accepted 23/06/2004}

   \abstract{
Spherically symmetric hydrodynamical outflows accelerated thermally in the 
vicinity of a compact object are studied by generalizing an equation of state 
with a variable effective polytropic index, appropriate to describe 
relativistic temperatures close to the central object and nonrelativistic ones
further away. Relativistic effects introduced by the Schwarzschild metric and 
the presence of relativistic temperatures in the corona are compared with 
previous results for a constant effective polytropic index and also with 
results of the classical wind theory. 
By a parametric study of the polytropic index and the location of the sonic 
transition it is found that 
space time curvature and relativistic temperatures tend to increase the 
efficiency of thermal driving in accelerating the outflow.  
 Thus conversely to the classical Parker wind, the outflow is 
accelerated even for polytropic indices higher than $3/2$.
The results of this simple but fully relativistic extension of the 
polytropic equation of state may be useful in simulations of outflows 
from hot coronae in black hole magnetospheres.}

\maketitle


\section{Introduction}

Relativistic outflows are commonly inferred from observations of collimated 
winds (jets) in Galactic x-ray binaries and supermassive black holes in 
active galactic nuclei and quasars (\cite{Birettaetal99}; Livio 2002). 
Also, observations infer coronae with rather high temperatures in microquasars 
(e.g., \cite{Corbeletal03}) and AGN  (e.g., \cite{RozanskaCzerny00} and refs. 
therein). Although there are still some ambiguities in the interpretation of 
these observations, usually temperatures up to $10^9$K for electrons/positrons 
and $10^{12}$K for protons are usually inferred. 
Nevertheless, even if such temperatures were to correspond to closed field 
line regions, by using the analogy with the solar corona, one may extrapolate 
that such high temperatures could also exist in the open field line regions 
where the outflow is 
accelerated. The heating mechanism to obtain these high temperatures can be of 
magnetic origin (\cite{HeinzBegelman00}). Alternatively, accreted material 
falling onto the central black hole may be decelerated via a shock, feeding 
the 
magnetosphere with a hot plasma. Discussions on the possibility of such shock 
formation in an accretion disk that leads to a heated corona have been 
already given in the literature in the so called CENBOL model
(\cite{Chakrabartietal96}, \cite{Chattopadhyayetal04}, Das, 2000, and
references therein). These papers have also discussed how radiation
and coupling with the photon distribution has to be taken into account 
properly.

The inevitable result of such hot atmospheres is that they expand 
supersonically and at large distances thermal energy is converted to bulk flow 
kinetic energy. For example, \cite{Ferrarietal85}, \cite{Das00} and 
Chattopadhyay et al. (2004) have 
suggested that  outflows  can be thermally accelerated to relativistic 
speeds. 
However, such treatments assumed the classical polytropic equation of state 
which prevents from studying consistently the relativistic temperatures in the 
corona and the nonrelativistic ones farther away. 
The first effort to use an equation of state appropriate for outflows 
containing both ultrarelativistic and classical temperatures has been used only
for studying spherical accretion flows and only in the adiabatic case
(\cite{Mathews71}; \cite{BlumenthalMathews76}). 
Moreover, Blumenthal \& Mathews (1976) discussed
  the topology of the Mach number variation in the solution 
of an adiabatic wind.

  However, as it is well known (\cite{Parker60}) the topology 
of the Mach number does not give us information on whether the flow itself 
is accelerated or not.

In this paper we extend the equation of state used in 
Blumenthal \& Mathews (1976) for 
nonadiabatic flows and our goal is to investigate the efficiency of thermal 
driving under extreme relativistic conditions. 
For example, in the classical polytropic solar wind theory, the 
polytropic index needs to be smaller than $3/2$ in order to obtain an 
accelerated wind solution (\cite{Parker60}). This means that a minimal 
extension of the corona is required. Here we explore how this value is changed 
under relativistic conditions.  In order to simplify the study, we shall focus 
our attention on spherical, steady  and radial hydrodynamic outflows at large 
distances from a stationary compact object. Also for simplicity we shall use 
Schwarzschild's metric.     

 Furthermore, we will not discuss how the
pressure and the heating may include coupling with the radiation field, 
especially in the case of pair production, as our 
goal is to keep the discussion at a basic level and this has already
been done elsewhere (e.g., \cite{Chattopadhyayetal04}).

In the next section we outline the governing equations with particular emphasis
to an equation of state appropriate to relativistic temperatures close to the 
base of the outflow and classical ones further away. In Sec.~3 we derive for a 
given asymptotic value of the polytropic index the range of allowed 
locations of the sonic surface and also the limits of the asymptotic
speed. In Sec.~4 via a parametric study we present a comparison of our 
model
with the nonrelativistic Parker wind and relativistic winds with constant
polytropic index. An astrophysical application 
is outlined in Sec.~5 and the results are summarized in Sec.~6.    

\section{Basic Equations}

The flow of a relativistic fluid is governed by conservation of the number 
of particles and energy-momentum,
\begin{eqnarray}
\label{4continuity}(n u^a)_{;a}  = 0 \,,\\
\label{Tenseur}(T^{a b})_{;b} = 0 \quad {\rm with} 
\quad T^{a b}= (n w/c^2) u^{a} u^{b}+ P g^{a b} \,, 
\end{eqnarray}
where $n$ is the proper number density of particles 
(electrons/protons, or, electrons/positrons) 
in the comoving frame of the fluid and 
$u^a=(\gamma c ;  \gamma \vec{v}) $
is the  fluid four-velocity, 
with $\vec{v}$ the spatial velocity and $\gamma
=\left(1-{v}^2/c^2\right)^{-1/2}$ the Lorentz factor.

The proper enthalpy per particle, $w$, is the sum  
of the proper internal energy per particle (including the proper mass), $e$,
and the proper pressure $P$ divided by $n$,
\begin{equation}
w = e + \frac{P}{n} \label{enthalpy}
\,.
\end{equation} 

The space-time around the central compact object
is described by using Schwarzschild's metric wherein 
in order to treat analytically the fluid equations the simplest 
way is to express all quantities in a 3+1 split space-time, 
\begin{equation}
{d}s^2 = -\alpha^2 c^2 {d}t^2 + \frac{1}{\alpha^2} 
{d}r^2+r^2 {d}\theta^2+r^2  
\sin{\theta}^2 {d}\phi^2
\,,
\end{equation}
where 
\begin{equation}
\alpha = \sqrt{1 - \frac{2 {\cal G} {\cal M}_{\star}}{c^2 r}}
= \sqrt{1-\frac{r_{\rm G}}{r}}
\,,
\end{equation}  
is the lapse or redshift factor induced by gravity at a distance $r$ from the 
central object (e.g., a black hole) of mass ${\cal M}_{\star}$, 
written also in terms of the Schwarzschild or gravitational radius 
$r_{\rm G}$,
\begin{equation}
r_{\rm G}=\frac{2{\cal G}{\cal M}_{\star}}{c^2}
\,.
\end{equation} 

Assuming spherical symmetry such that the velocity is purely radial, 
$\vec{v}=v \hat r$, 
Euler's equation reduces to the following single differential 
equation, 
\begin{equation}
\frac{1}{2}\gamma^2 n \frac{w}{c^2} \frac{{d} {v}^2}{{d} r} 
+ n \frac{w}{c^2}c^{2}\frac{{d} \ln \alpha }{{d} r} +
 \frac{{d} P}{{d} r} =0
\,.
\label{euler1}
\end{equation}

By combining Eqs.  (\ref{Tenseur}) and (\ref{enthalpy}), we
get that the variations of the specific enthalpy and internal 
energy are
\begin{equation}
{d}w=\frac{1}{n}{d}P\,,\quad
 {d}e=\frac{P}{n^2}{d}n \label{E_generalise}
\,.
\end{equation}

Similarly, the  conservation of the number of particles for a spherical 
wind, Eq.  (\ref{4continuity}), takes the following simpler form, 
\begin{equation}
\frac{d}{{d}r}(\alpha \gamma n v r^2) = 0
\,.
\label{continuity0}
\end{equation}

In the following, we shall adopt an equation of state appropriate to describe 
relativistic temperatures close to the central object and nonrelativistic ones 
further away and also derive its ultrarelativistic and classical limits. 

\subsection{Closure of the fluid equations}

To close the system of equations, (\ref{euler1}-\ref{continuity0}) we 
should specify an equation of state that
relates the entropy per particle $s=s(w,P)=s(e,P)$ to the pressure, $P$, and 
the enthalpy per particle, $w$, or the internal energy per particle, $e$. 
In the literature (e.g., \cite{Michel72}; \cite{Chakrabartietal96}; \cite{Das99}) 
various authors have adopted a 
classical adiabatic or polytropic equation to replace the equation  of state
\begin{equation}\label{classicalpolytrop}
P  \propto  n^\Gamma
\,,
\end{equation} 
where $\Gamma$ is a constant. However, in this form $\Gamma$ should attain 
different values whether the particles are in a relativistic or a classical 
thermodynamical regime. It is well known that for a monoatomic gas, 
in adiabatic flow the index is $5/3$ in the classical regime 
with nonrelativistic temperatures ($k_{\rm B} T \ll mc^2$),
and $4/3$ in the ultra-relativistic one ($k_{\rm B} T \gg mc^2$). 
(Here $k_{\rm B}$ is the Boltzmann constant
and $mc^2$ is the rest mass energy per particle.)
Thus we can define an effective polytropic index
\begin{equation}\label{Gammaeff}
\Gamma_{\rm eff} = \frac{{d}\ln{P}}{{d}\ln{n}} 
\end{equation}
which should vary with temperature. In other words, the previous equation of 
state with a constant index cannot account 
for media with a transition from ultrarelativistic to nonrelativistic 
temperatures, as inferred from observations of coronae in AGNs.

The scalar isotropic pressure of a single perfect fluid is given by 
(\cite{Mathews71}; \cite{Synge57})
\begin{equation}\label{Synge}
P=\frac{1}{3} n \epsilon_{0}
\left(\frac{e}{\epsilon_{0}} -\frac{\epsilon_{0}}{e} \right)
\,,
\end{equation}
where $\epsilon_{0} = mc^{2}$. Its validity for collisional and 
collisionless fluids is discussed in detail in 
Blumenthal \& Mathews (1976) and we refer the reader to 
section III of their paper.

By integrating Eqs. (\ref{E_generalise}) and (\ref{Synge})  we obtain 
a generalized equation of state for an adiabatic (ideal) fluid 
(see \cite{BlumenthalMathews76}, Eq. 3.3), 
\begin{equation}\label{monoatomic}
e^2 - \epsilon_{0}^2 
= \kappa n^{\frac{2}{3}}
= \kappa n^{\Gamma_{\rm ad} -1}
\,,
\end{equation}
which represents the equation of state for an adiabatic flow. Here 
$\Gamma_{\rm ad}=5/3$ is the real adiabatic index related to the number of 
degrees of freedom of the particles.

For studying stellar interiors, a classical trick to study different equations 
of state and their hardness is to use the same equation,  called 
polytropic equation of state, with a
different value of the adiabatic index, assuming the number of degrees of 
freedom of the particles has changed. In this case $\Gamma_{\rm ad}\ne 5/3$, 
but the system still evolves adiabatically.
E. Parker to study the solar wind proposed (\cite{Parker60})
to follow the same path using Eq. \ref{classicalpolytrop}, 
assuming $\Gamma$ being any value 
between 1 and the adiabatic value $5/3$ to mimic the coronal heating in a 
simple but hidden way. As explained in \cite{STT99} this is equivalent to
use a generalized enthalpy and internal energy which includes the heating.

Following Parker's initial approach of the polytropic solar wind, we assume 
also here that the enthalpy and the particle density are related  by Eq. 
(\ref{E_generalise}) but
relating $e$ to the particle density via
\begin{equation}
e^2 - \epsilon_{0}^2
=\kappa n^{\Gamma -1}
\label{defpolytrope}
\,,
\end{equation}
with  again a value of the constant polytropic index $\Gamma$ which can be 
anything between 1 (isothermal) and 5/3 (adiabatic value), or higher in case of
energy losses. The basic idea is to 
generalize to heated relativistic winds the approach of 
Blumenthal \& Mathews (1976) for relativistic accretion flows like Parker
did with Bondi's accretion models. Of course at this stage $w$ (resp. $e$) is
a generalized specific enthalpy (resp. generalized specific internal energy)
which includes the energy brought via heating from the external
medium. We recall in Appendix \ref{appendix} how to calculate the extra 
heating implicitly in the case of a monoatomic ionized plasma.

By combining Eq. (\ref{defpolytrope})  and (\ref{E_generalise}) we get a 
generalized relation between pressure and density which includes the proper 
mass energy per particle, $\epsilon_{0}=mc^2$ ,
\begin{equation}\label{polytropePrho}
P =  \frac{\Gamma -1}{2}\frac{\kappa n^{\Gamma}}
{\sqrt{\epsilon_{0}^2 + \kappa n^{\Gamma - 1}}}
\end{equation}
for $\Gamma \neq 1$. Note that Eq. (\ref{polytropePrho}) is a generalization 
of the usual form of the so-called equation of state with the inclusion of 
relativistic thermal effects. Thus the effective polytropic index 
(Eq. \ref{Gammaeff})
takes the form
\begin{equation}
\Gamma_{\rm eff} = \Gamma - \frac{\Gamma-1}{2} 
\frac{\rho}{1 + \rho}
\,,
\end{equation}
where we have introduced a new dimensionless function $\rho$ 
\begin{equation}
\rho =\frac{e^2 - \epsilon_{0}^2}{\epsilon_{0}^2} 
= \frac{\kappa n^{\Gamma - 1}}{\epsilon_{0}^2}
\,.
\end{equation}
All thermal effects appear in this single function $\rho$, hence decreasing 
the number of free parameters in the model. 

Finally, the local sound speed $v_s= \beta_{\rm s} c $ can be written as 
a function of $\rho$,
\begin{eqnarray}
\beta_{\rm s}^2 &\;=\;& \frac{v_{\rm s}^2}{c^2} 
= \frac{1}{w} \frac{{d} P}{{d} n}\,\nonumber\\
&\;=\;&\frac{\Gamma - 1}{2} 
\left\{1-\frac{2+(3-\Gamma)\rho}{(1+\rho)[2+(\Gamma+1)\rho]}\right\}
\label{betasrho}
\,.
\end{eqnarray}
The logarithmic derivative of the sound speed with density is
\begin{equation}
\frac{d \ln \beta_{\rm s}^2}{d \ln \rho}=
\frac{(\Gamma+1)(3-\Gamma) \rho^2 + 4 (\Gamma+1)  \rho + 4 \Gamma}
{( \rho +1) \left[2 \Gamma + (\Gamma+1)  \rho \right] 
\left[ 2 + (\Gamma+1)  \rho \right] }
\,,
\label{derivative_sound}
\end{equation}
and, for $\Gamma < 3$, it is positive.
Thus, with $\Gamma < 3$ it follows that 
\begin{equation}\label{betamax}
\beta_{\rm s}^2 \le \frac{\Gamma-1}{2}
\,,
\end{equation}
with the equality holding in the high temperature limit 
($\rho\longrightarrow \infty$). 
Note that for $\Gamma<5/3$, we get the usual condition
$\beta_{\rm s}^2 < 1/3$ (\cite{Michel72}).

\subsection{Nonrelativistic and ultrarelativistic limits}

In the limit of a nonrelativistic fluid, $\epsilon_{\rm th}\ll\epsilon_{0}$, 
we may calculate the thermal energy per particle or random thermal energy 
(\cite{Mathews71}), 
\begin{eqnarray}
\epsilon_{\rm th} & = e - \epsilon_{0}
=\sqrt{\kappa n^{\Gamma -1} + \epsilon_{0}^2} - \epsilon_{0}
\nonumber\\
&\approx
\frac{\displaystyle \kappa }{\displaystyle 2\epsilon_{0}}n^{\Gamma -1}
{{\rm,\qquad\qquad if\ }\epsilon_{\rm th}\ll\epsilon_{0}}
\,.
\end{eqnarray}
In such a case, $\rho$ is twice the ratio of the thermal 
energy per particle and the mass energy
\begin{equation}
\rho \approx 2\frac{\epsilon_{\rm th}}{\epsilon_{0}} 
\,.
\end{equation}
For classical temperatures, wherein the thermal energy is negligible 
compared to the mass energy, $\epsilon_{\rm th}\ll \epsilon_0$,  we 
have that $ \rho \ll 1$ and  
\begin{equation}
P \simeq  \left(\Gamma-1\right)n\epsilon_{\rm th}
\simeq \; \left(\Gamma-1\right)\frac{\kappa}{2\epsilon_0} n^{\Gamma}
\,,
\end{equation}
such that the effective polytropic index is exactly $\Gamma_{\rm eff}=\Gamma$.
In general, as temperature decreases with distance, 
$\Gamma$ always represents the asymptotic value of the 
effective polytropic index. This allows us to have a simple albeit artificial
way to model relativistic coronae as in the classical Parker wind.

On the other hand, in the ultrarelativistic domain where the thermal energy 
is much larger than the mass energy ($\epsilon_{\rm th} \gg \epsilon_0 $ 
$\Leftrightarrow$ 
$ \rho \gg 1$) we have
\begin{equation}
P \simeq  \frac{\Gamma-1}{2}
{\kappa^{\frac{1}{2}} 
n^{\frac{\Gamma + 1}{2}}}
\,.
\end{equation}
The effective polytropic index becomes now $\Gamma_{\rm eff}=(\Gamma+1)/2$.

\begin{figure}
\rotatebox{0}{\includegraphics[width=8cm]{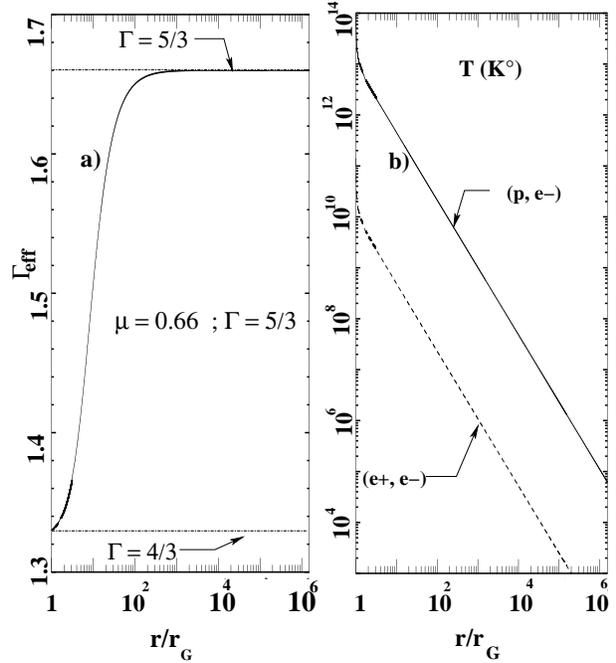}}
\caption{(a) 
Plot of the effective polytropic index $\Gamma_{\rm eff}$ of an adiabatic 
outflow showing the transition of $\Gamma_{\rm eff}$ from the ultrarelativistic
value of $4/3$ to the classical one $5/3$. 
The corresponding temperature profile is plotted in (b) 
both, for electron/proton [$k_{\rm B}T=P/(2n)$] and pair plasmas
[$k_{\rm B}T=P/n$] and has ultrarelativistic temperatures at the base 
which decrease to classical ones further out.
}
\label{Fig1}
\end{figure}

The effective polytropic index and corresponding temperature for an adiabatic 
flow with relativistic temperatures at the base and classical behavior further 
away are plotted in Fig. \ref{Fig1}. 
These solutions are discussed in more detail later on. We simply note here  
that for an adiabatic flow, the polytropic index increases from $4/3$ at the 
base where the thermal energy equals or exceeds the mass energy (i.e. 
$T\ge 10^{12}K$ for protons, or, $T\ge 10^{9}K$ for positrons) to $5/3$ 
asymptotically where the temperature is much lower (i.e., $T\ll 10^{12}K$ for 
protons or $T\ll 10^{9}K$ 
for positrons), as expected. To emphasize this property of the equation
of state, we have plotted in Fig. \ref{Fig1} a case with ultrarelativistic 
temperatures at the
base. Note that the coronal temperature close to the compact object are very 
sensitive to the value of the parameter $\mu$, as discussed in the following.

\section{Equations in dimensionless form and parameters}

\subsection{Momentum and continuity equations}

In the following all quantities are defined in dimensionless form. First, 
velocities are normalized to the speed of light $\beta = v/c$ such that 
$\beta_{\rm s} = v_{\rm s}/c$ is the (variable) dimensionless sound 
speed. Second, distances could have been normalized at the  gravitational 
radius
$r_{\rm G}$. However, we found easier to normalize all quantities at the sonic 
surface $r_\star$, such that we can define a dimensionless radius,
\begin{equation}
R=r/r_\star
\,.
\end{equation}
A crucial  parameter of our model is
\begin{equation}
\mu=\frac{r_{\rm G}}{r_\star} 
= \frac{v_{\rm esc}^2(r_{\star})}{c^2}
\,,
\end{equation}
which is the square of the escape speed at the sonic radius normalized 
to the speed of light. It is also 
the ratio of the gravitational radius to the
sonic distance. Thus $\mu$  measures the strength of the gravitational field 
and the distance between the central black hole and the sonic 
surface.\footnote{This parameter $\mu$ is similar to the parameter $m$ used by 
Daigne \& Drenkhahn (2002) except that we use the sonic surface while they use 
the Alfv\'en radius for normalization.}

The Euler equation and the continuity equation can be written in dimensionless 
forms,
\begin{equation}
\frac{\beta^2 - \beta_s^2}{1-\beta^2} 
\frac{{d} \ln{\beta^2}}{{d} R}
=\frac{4\beta_{\rm s}^2} {R}
-\frac{1-\beta_{\rm s}^2}{R} \frac{\mu}{R-\mu}
\label{euler}
\,.
\end{equation}

\begin{equation}
\frac{\beta^2 - \beta_s^2}{\beta^2}
\frac{{d} \ln{\rho}}{{d} R}=
(\Gamma-1)\left[-\frac{2}{R}+\frac{1-\beta^2}{\beta^2} 
\frac{1}{2 R}\frac{\mu}{R-\mu}\right]
\label{continuity2}
\,.
\end{equation}

In a nonrelativistic wind, the radius of the star is
independently given from its mass. Conversely if the central object is a black
 hole, the gravitational radius provides a natural length scale. However, the 
wind cannot start and accelerate at this particular distance. In fact, it has 
to start at a radius $r_o$ located obviously above the Schwarzschild radius in 
the corona. In other words the wind solution should start at a dimensionless
radius $R_o>\mu$ which will be an output of the integration starting at the 
sonic surface and integrating up wind. 

\subsection{Integration of equations and range of values of $\mu$}

This model has two free parameters
that affect the solution of the differential equations
(\ref{euler}), and (\ref{continuity2}).\footnote{The other
parameters are the mass of the central object ${\cal M_*}$,
the mass-loss rate $\dot{m}$, and the constant $\kappa$.} 
The first parameter, $\Gamma$, is 
the value of the effective polytropic index at infinity. The second parameter,
$\mu$, is the ratio of the Schwarzschild's radius to the radius at the
 sonic  surface.

Eqs. (\ref{euler}) and (\ref{continuity2}) have a singularity corresponding to 
the sonic surface, which we may call hydrodynamic horizon by analogy with the 
black hole horizon. At the  sonic  surface
 
the right hand sides of the two equations 
vanish because the left hand sides vanish.

For a detailed discussion of the nature and position
of the critical surfaces in astrophysical winds, see  
\cite{Tsinganosetal96}.

This gives the following 
criticality condition,
\begin{equation}\label{betastar}
\beta_{\rm s \star}^2 = \frac{\mu}{4 - 3 \mu}
\,.
\end{equation}
Note that the value of $\rho$ at the sonic surface, $\rho_\star$,  
follows from Eq.(\ref{betasrho}).  

Since $\beta_{\rm s \star}^2<1$,  
the above criticality condition requires that we should always have $\mu<1$.  
Furthermore, by combining Eqs. (\ref{betamax}) and  (\ref{betastar}) 
we find an upper value for $\mu$
\begin{equation}\label{mu_max}
\mu <\mu_{\max} =\frac{4 ( \Gamma - 1 )}{3 \Gamma -1}
\,.
\end{equation}

Once we get the value of 
$\beta_{\rm s \star}$ ($\Leftrightarrow\rho_{\rm s \star}$)
from the sonic criticality condition, we deduce the values of 
the slope of the velocity at the critical surface using De L'Hopital's rule
in Eqs. (\ref{euler}) and (\ref{continuity2}) which after some lengthy 
calculations gives
\begin{equation}\label{slopestar1}
A{\left.\left[\frac{{d} \beta^2}{{d} R}\right]^2\right|_{\star}} 
+ B{\left.\frac{{d} 
\beta^2}{{d} R}\right|_{\star}} + C = 0
\end{equation}
\begin{eqnarray}
A =   1 + (\Gamma - 1) \left.\frac{{d}\ln{\beta^2_{\rm s}}}{{d}\ln{\rho}}
\right|_{\star}
\frac{(4 - 3 \mu)}{8( 1 - \mu )}
\,, \\
B = 
\left.\frac{{d}\ln{\beta^2_{\rm s}}}{{d}\ln{\rho}}\right|_{\star}
(\Gamma - 1)\frac{\mu}{1 - \mu}
\,, \\
C= \frac{-16\mu^2}{(4 - 3 \mu)^3}\left(1 - 
\left.\frac{{d}\ln{\beta^2_{\rm s}}}{{d}\ln{\rho}}\right|_{\star}( \Gamma - 1)
\frac{(4 - 3 \mu)^2}{8 (1 - \mu)}\right)
\,.
\label{C}
\end{eqnarray}

\begin{figure}[h] 
\rotatebox{0}{\includegraphics[width=8cm]{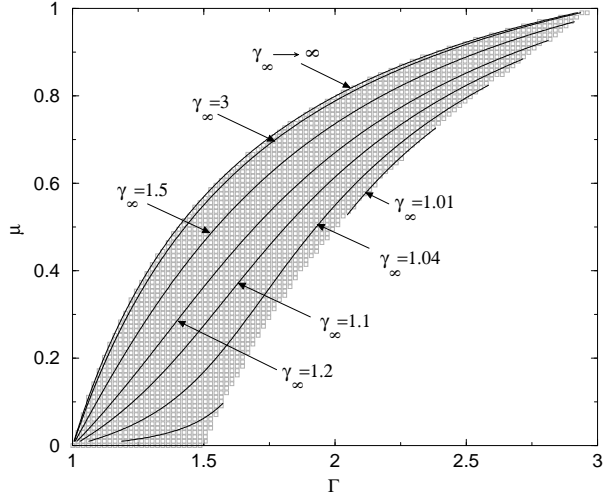}}
\caption{
Contour plots of constant terminal Lorentz factor $\gamma_\infty$ are shown 
within 
the limiting values of $\mu$, $\mu_{\min}$ and $\mu_{\max}$  versus $\Gamma$, 
with the  grey filled region showing the allowed wind space parameter. 
Note that $\mu_{\max}$ corresponds to $\gamma_\infty \longrightarrow \infty$.  
  }
\label{Fig2}
\end{figure}

As we are interested for wind type solutions, the derivative 
$\left. d \beta^2/dR \right|_\star$
at the sonic point should be positive. 
By noting that $A\,, B>0$, this is equivalent to the condition $C<0$.
Then, from Eq. (\ref{C}) and employing eqs. (\ref{betasrho}), 
(\ref{derivative_sound}), and (\ref{betastar}) we get an inequality for $\mu $.
We solved this numerically and found the condition
$\mu>\mu_{\min}$ where the value of $\mu_{\min}$ as a function 
of $\Gamma$ is shown in Fig. \ref{Fig2} as the lower limit of the
grey filled region.

Thus, the criticality conditions for wind-type solutions reduces the 
acceptable values of $\mu$ to a limited range:
\begin{equation}
\mu_{\min} < \mu < \mu_{\max}
\,.
\end{equation}
\begin{itemize}
\item $\mu_{\max} = r_{\rm G}/r_{\star, \min}$ (see Fig. \ref{Fig2}) 
corresponds to the minimum distance between the Mach surface and the black 
hole, 
$r_{\star, \min}$, for which Eq. (\ref{betasrho})  
has an acceptable solution $\rho_\star=\rho_\star(\beta_{\rm s \star})$.   
For the limiting case $r_{\star} \longrightarrow  r_{\star, \min}$, 
$\rho_\star \longrightarrow  \infty$. 
If the sonic surface is too close to the black hole, gravity
becomes so important that the critical point disappears. 

\item $\mu_{\min} = r_{\rm G}/r_{\star, \max}$ (see Fig. \ref{Fig2}) 
corresponds to the maximum 
distance,  $r_{\star, \max}$, between the 
 sonic surface and the 
black hole above which the two critical solutions are accretion-type 
solutions. 
\end{itemize}

{}From Fig. \ref{Fig2} we see that there is a maximum value,  $\Gamma=3$, 
beyond which $\mu_{\min} > \mu_{\max}$,
because the cooling is too strong to have a wind type solution 
similarly to Begelman (1978).

\subsection{Bernoulli constant}

The conservation of the number of particle can be integrated to give a first 
constant of the system, which corresponds to the mass loss rate or the mass 
flux $\dot m$
\begin{equation}
\alpha \gamma n v r^2 = \dot m
\,,
\label{continuity}
\end{equation}

Similarly, by integrating Euler's equation (\ref{euler1}), 
we obtain the Bernoulli equation
$E=\alpha \gamma w$, which is given by
\begin{equation}
E=\epsilon_{0} \left(1-\frac{\mu}{R}\right)^{1/2} 
\frac{1}{\sqrt{1-\beta^2}}
\frac{1+\left(1+\Gamma\right)\rho/2}{\sqrt{1+\rho}}
\,.
\end{equation}
Note that asymptotically ($R\rightarrow \infty$,
$\rho \rightarrow 0$) the latter equation implies that the
Lorentz factor is $\gamma_\infty = E/ \epsilon_{0}$.
On the other hand, by using the values of $\beta_\star=\beta_{\rm s \star}$ 
and $\rho_\star$ at the sonic  surface

(eqs. [\ref{betastar}] and [\ref{betasrho}]),
we get the Bernoulli constant $\gamma_\infty = E/ \epsilon_{0}$
as a function of $\Gamma $ and $\mu$, see Fig. \ref{Fig2}.

Instead of integrating the differential equations which is the way we followed
after Parker (1960), we could also plot the isocontours of constant $E$ and 
$\dot m$. The two methods are equivalent. We have checked that 
while integrating the differential equations $E$ remained constant. 

\section{Results and comparison with the nonrelativistic cases}

In the following we present the results of our parametric study for various 
values of the polytropic index $\Gamma$ and the gravitational parameter $\mu$. 
We also compare them with the corresponding results in the analysis of 
Blumenthal \& Mathews (1976) for an adiabatic polytropic index $\Gamma = 5/3$ 
and studies using a classical equation of state with a constant polytropic 
index $\Gamma$.    

\subsection{Parametric study}

We solved numerically the equations by using a Runge-Kutta scheme starting at 
the sonic  surface

 and integrating both upstream towards the black hole and 
downstream to infinity.  

\begin{figure*}[t]
\rotatebox{180}{\includegraphics[width=6cm,height = 6cm]
{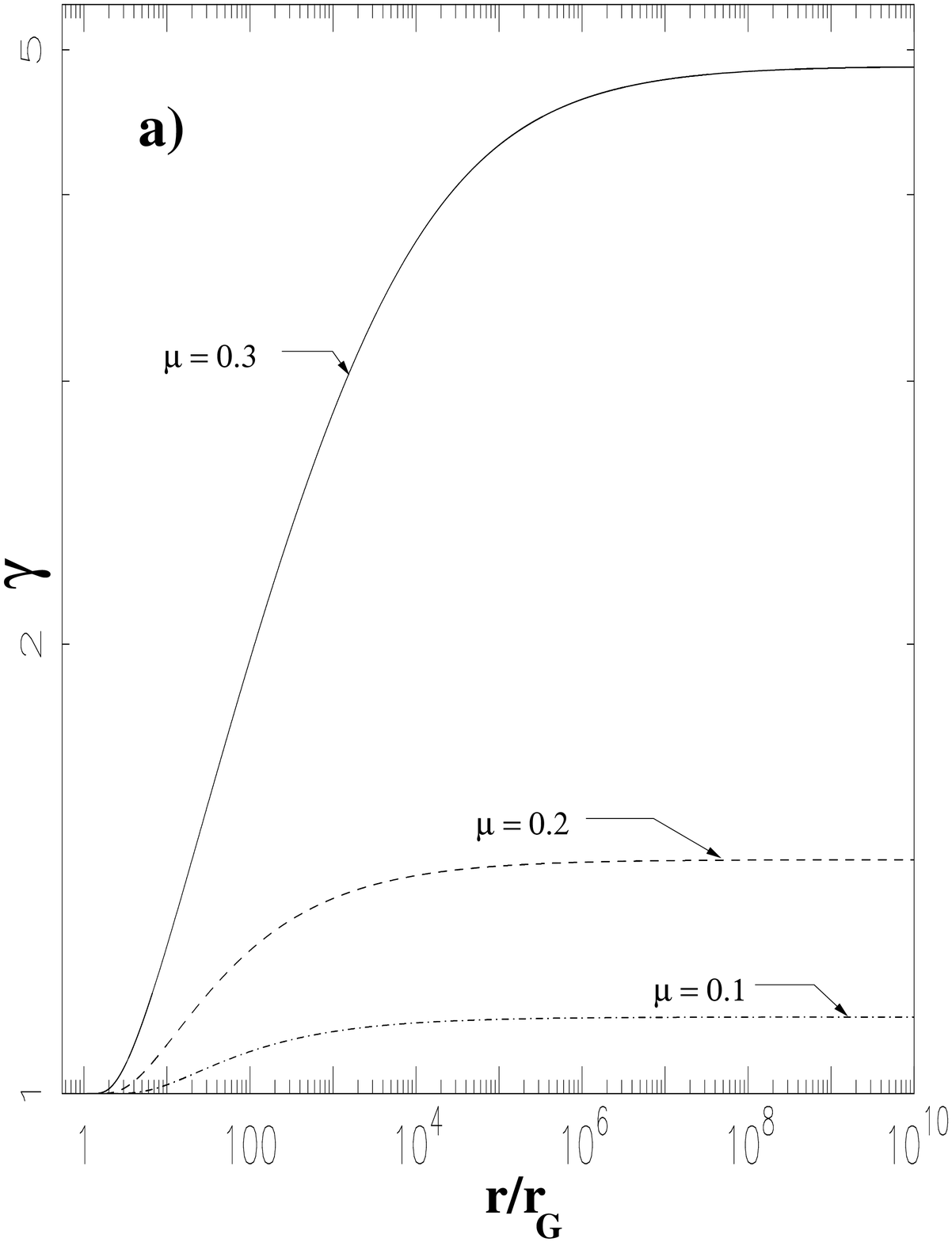}}\rotatebox{270}{\includegraphics[width=6cm,height = 6cm]
{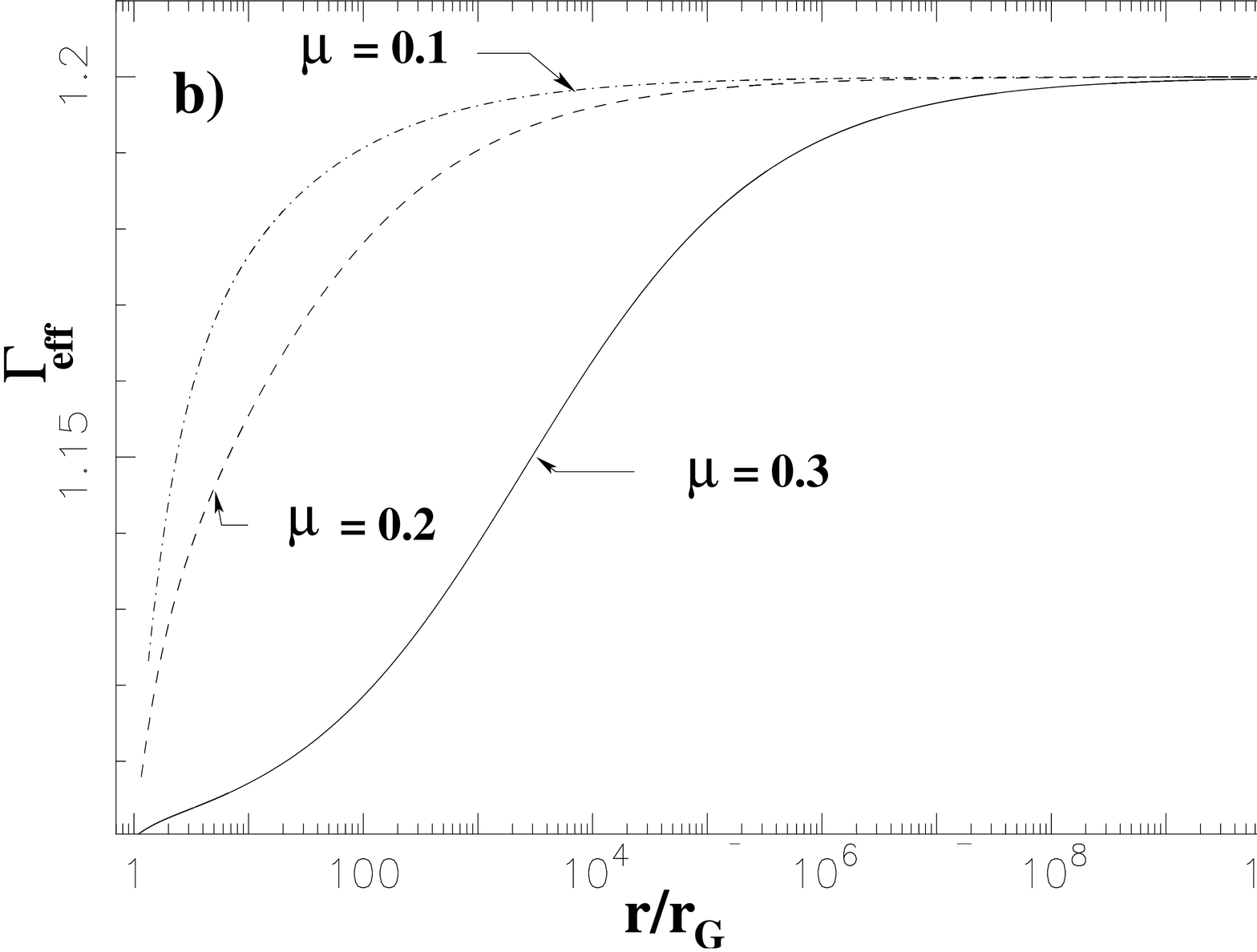}}\rotatebox{180}{\includegraphics[width=6cm,height = 6cm]
{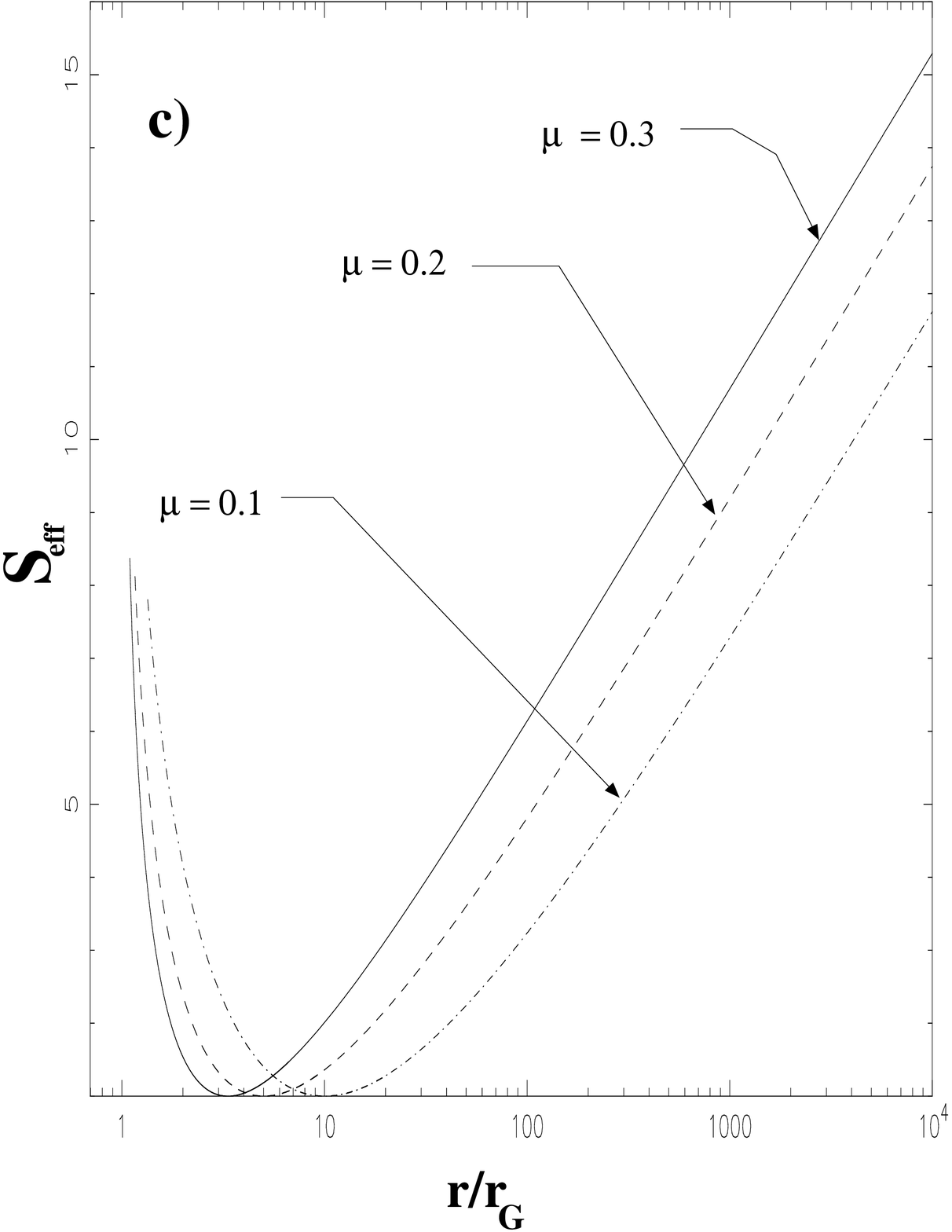}}
\caption{
For $\Gamma = 1.2$ we show respectively as functions of the radial distance, 
the Lorentz 
factor in (a), the effective polytropic index  in (b) and the effective 
cross-section of the equivalent De Laval Nozzle in (c) for various wind 
solutions corresponding to three values of $\mu$ labeling each curve.
}
\label{Fig3}
\end{figure*}

The behavior of the solutions for various values of $\mu$ 
is displayed in Figs. \ref{Fig3}.
As $\mu$ increases the sonic  surface

 moves inwards closer to the compact object. 
Thus, the supersonic region of the outflow exposed to thermal energy 
deposition 
is enlarged. Consequently the Bernoulli energy gets larger for a given 
gravitational energy.
This induces a better conversion of thermal energy into kinetic energy.
In other words, as $\mu$ increases the distance between the corona and the 
black hole decreases and, to support gravity, the pressure becomes higher.
 The pressure driven wind passes through the 
critical point faster and 
subsequently transforms into a relativistic wind at large distances from the 
black hole. The situation is similar to that of nonrelativistic winds with 
momentum and heat addition (\cite{LeerHolzer80}) wherein only for 
heat/momentum addition in the supersonic part of the flow the terminal wind 
speed increases. The mass loss is unaffected in such a case and is only 
affected for momentum/heat addition in the subsonic region. 
As illustrated by Fig. \ref{Fig3}b, the higher is $\mu$, the more extended 
downstream is the area of the flow where the effective polytropic index is 
smaller than $\Gamma = 1.2$ and thus the wind is more efficiently accelerated.  
This more efficient conversion of thermal energy for higher values of $\mu$
maybe also illustrated by calculating the cross-section of the equivalent De 
Laval Nozzle (\cite{Parker60}). 
\begin{equation}\label{cross-section}
\frac{{\rm d}\ln S_{\rm eff} }{{\rm d} R}
= \frac{2}{R}-\frac{1\,-\,\beta_{s}^2}{2\,R\,(R-\mu)}\;\frac{\mu}{\beta_{s}^2}
\end{equation}
where $S_{\rm eff}$ is the effective cross-section.
As shown in Fig. \ref{Fig3}c, the larger is $\mu$,  the 
larger is the opening of the effective cross-section. 

\subsection{Comparison with the classical wind}

In the classical limit, the horizon goes to the center of the 
gravitational well, $r_G \rightarrow 0$ and consequently
$\mu \rightarrow  0$. 
The index $\Gamma$ is the only free parameter which is left together with 
the mass loss rate.
 
Then, in the 
classical limit the condition at the  critical
point reduces to the usual expression, 
 \begin{equation}
\frac{\beta_{\rm s \star}^2}{\mu} = \frac{1}{4} 
\Leftrightarrow v_{\rm s\star}^2 = \frac{ {\cal G} 
{\cal M}_{\star} }{ 2 r_\star}
\,,
\end{equation}
where the sound speed at the sonic  surface

is equal to half of the escape 
speed.
 From this single relation, the distance between the stellar 
surface and the sonic surface is uniquely determined from the sound speed 
at the  sonic surface

 Conversely in the  relativistic case 
$\mu$ is a free parameter, 
so we can change the distance between the two surfaces 
and increase the effect of  thermal conversion in accelerating the wind.

\begin{center}
\begin{figure*}[t]
\rotatebox{270}{\includegraphics[
height=9cm]
{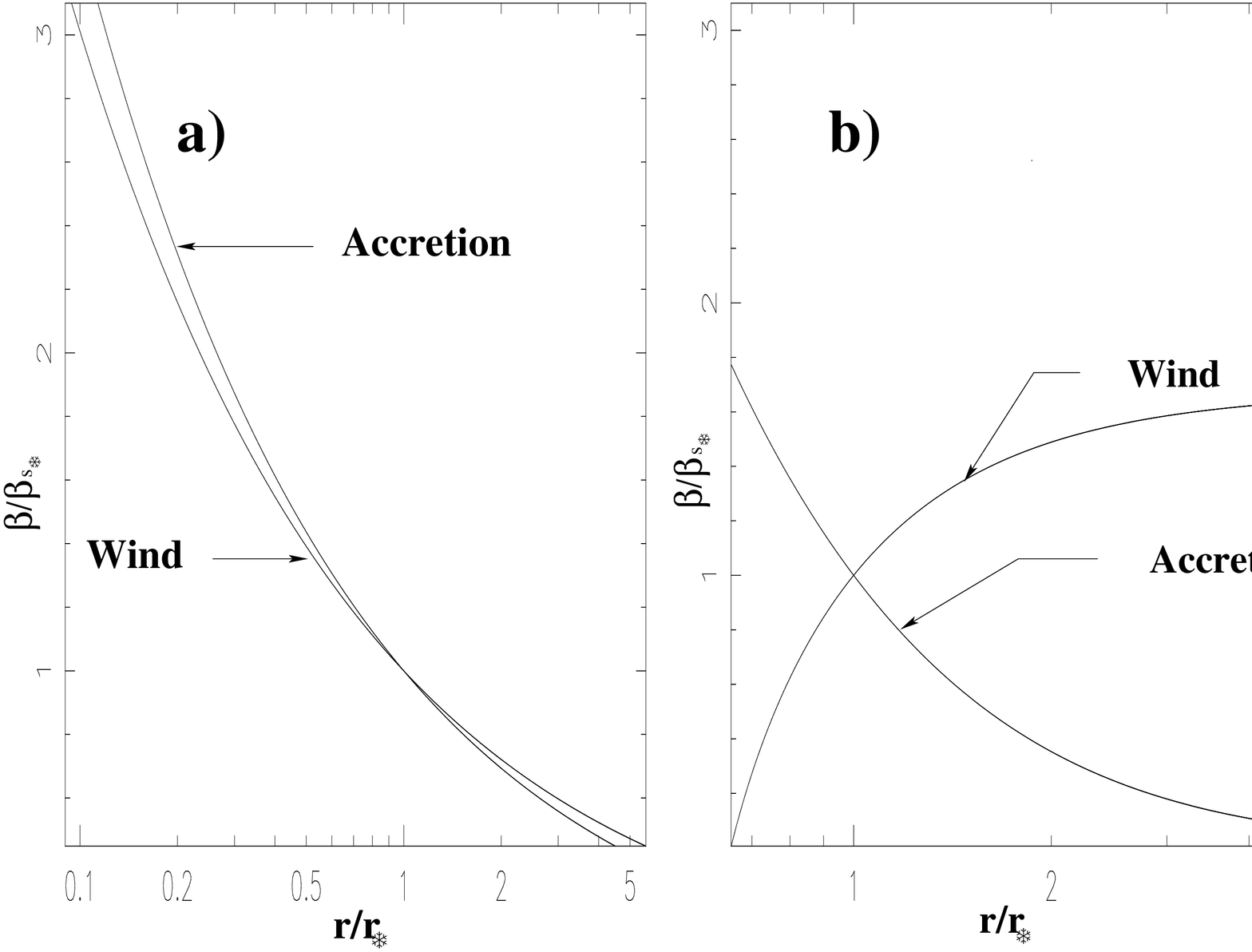}}\rotatebox{270}
{\includegraphics[
height = 9cm]{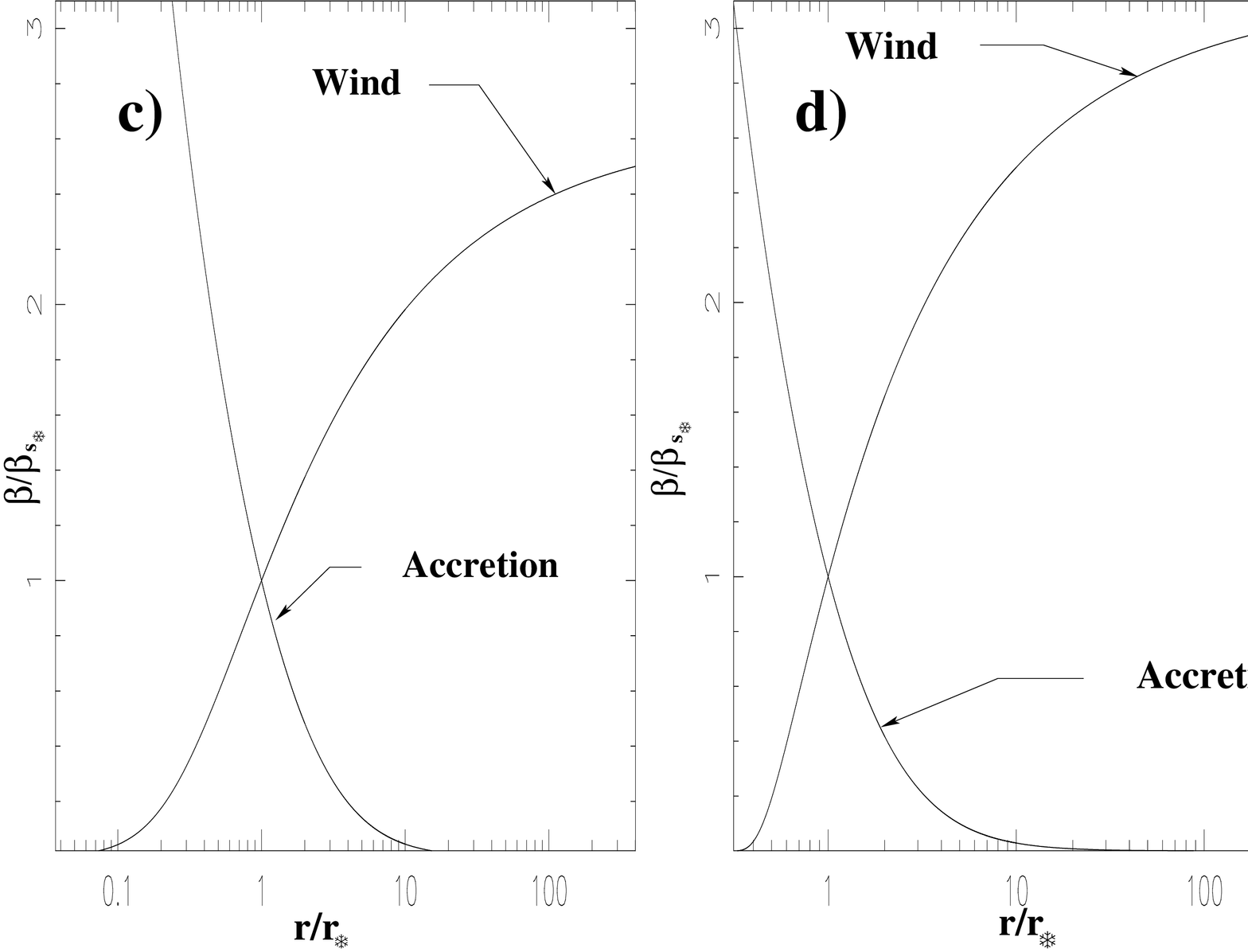}}
\caption{Comparison of the classical Parker solutions and the relativistic ones
for two sets of the parameters $\Gamma$ and $\mu$.
For an adiabatic polytropic index $\Gamma = 5/3$ and $\mu=0.65$ in a) is shown the nonaccelerating
classical Parker solutions while in b) the accelerating relativistic ones. 
For $\Gamma = 1.2$ and $\mu=0.2$ is shown in c) the classical Parker solutions and in d)
the relativistic ones. }
\label{Fig8}
\end{figure*}
\end{center}

For a classical polytropic wind there is no acceleration for 
$\Gamma > 3/2$ (see \cite{Parker60}) because then the acceleration at
the   sonic surface

becomes negative for both critical solutions and
no real wind-type solution exists. Conversely our relativistic solutions can
be accelerated even if $\Gamma > 3/2$ when $\mu$ increases. This is illustrated
in Figs. \ref{Fig8}a,b
where the topology of the adiabatic solution 
($\Gamma = 5/3$) is displayed in the
classical and the relativistic cases. In the first case both critical solutions
are decelerated while in the second one there is an accelerated wind type 
solution. This is also shown in Fig. \ref{Fig2}: for $\mu=0$ (classical limit,
along the horizontal axis),
 the highest value of $\Gamma$ to have an accelerated wind-type solution
is $1.5$. For non zero values of $\mu$ (relativistic regime), the domain of 
wind-type  solutions extends to higher values of $\Gamma$. 

The relativistic wind is characterized 
by the fact that from the compact object outwards the pressure decreases
faster. This is due to the presence of the redshift function $\alpha$ in the 
continuity equation (\ref{continuity}).  When matter flows away from the 
central object the proper volume of the fluid increases faster than the 
observed (spherical) volume because of $\alpha$,

which provokes the decrease of the proper density and consequently 
of the pressure. 
The larger pressure gradient leads to higher acceleration. This is equivalent 
to the well known acceleration due to overradial flaring in the
classical solar wind as studied in \cite{KoppHolzer76}.

In fact, expanding Euler's equation (\ref{euler}) to first order in the 
relativistic effects, i.e. for $\mu\ll 1$ and $\rho\ll 1$,  we find
\begin{eqnarray}
\frac{1}{2}  \frac{(\beta/\beta_{\rm s})^2-1}{1-\beta^2} 
\frac{\partial \ln{\beta^2}}{\partial R} =
\nonumber \\
\frac{2}{R^2}\left(R - \frac{\beta_{\rm s \star}^2}{\beta_{\rm s}^2}\right)
+  \frac{3 \mu }{2R^2} \frac{\beta_{\rm s \star}^2}{\beta_{\rm s}^2}
+  \frac{\Gamma \left(\Gamma -1\right) \rho}{R^2} 
\frac{\beta_{\rm s \star}^2}{\beta_{\rm s}^2}
\,.
\end{eqnarray}
The first term  of the right hand side
is the classical term. The second term
shows the effect of gravity in general relativity, related to the 
existence of a characteristic distance due to the metric of the black hole.  
This positive term increases the acceleration and the conversion of thermal 
energy into  kinetic energy. The third term
shows the influence of relativistic temperatures. It is also a positive term 
which 
increases the efficiency of the acceleration as long as the temperature 
remains relativistic.  These two terms are responsible for the acceleration 
obtained even for $\Gamma  > 3/2$.

Even for quasi-isothermal ($\Gamma \ga 1$)
flows we see that the relativistic winds are more rapidly accelerated than
their classical counter part (Fig. \ref{Fig8}c,d).

\subsection{Comparison  with a relativistic wind having a constant effective
polytropic index}

Let us consider next relativistic winds  with a {\it constant} effective polytropic. 
In this approach the same free parameters exist, 
$\mu$ and $\Gamma$. However for a relativistic polytropic wind where 
$\Gamma_{\rm eff}=\Gamma=$const, the expression of the sound speed and the 
Bernoulli equation are written as
\begin{equation}\label{betasonicclassic}
{\beta_{\rm s_c}}^2=\Gamma\frac{( \Gamma - 1 )\rho_c}{( \Gamma - 1 ) 
+\Gamma \rho_c}
\,, \quad \rho_c \equiv \frac{P}{n \epsilon_0} \,,
\end{equation}
\begin{equation}
E=\alpha \gamma\left( 1 +
 \frac{\Gamma}{\Gamma - 1} \rho_c \right) \epsilon_0
\,.
\end{equation}
The main difference in the behavior of the solutions with those using the
generalized equation of state appears in the zone close to the compact object 
where there is a transition between
classical and relativistic temperatures.

\begin{figure}[h]
\rotatebox{180}{\includegraphics[width=9cm,height = 8cm]{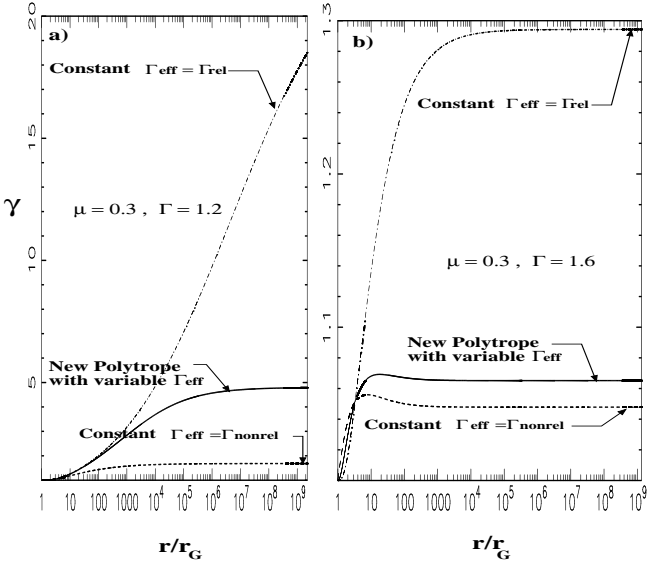}}
\caption{
Comparison between a variable and two constant $\Gamma$ polytropic 
equations of state for two solutions with $\mu=0.3$. 
In the left panel $\Gamma = 1.2$ while in the right panel $\Gamma = 1.6$.
In each panel, the two constant values $\Gamma_{\rm rel}=(\Gamma+1)/2$ and 
$\Gamma_{\rm nonrel} =\Gamma$ correspond to the limits of $\Gamma_{\rm eff}$ 
in the ultrarelativistic and the classical regimes, respectively.
}
\label{Fig9}
\end{figure}

For a fixed value of  $\mu$, we may compare the solutions of three 
equations of state describing a relativistic wind. {\it First}, by using the 
consistent equation of state with a {\underbar {variable}} 
$\Gamma_{\rm eff}$ and {\it second},  
for a {\underbar {constant}} polytropic index, either $\Gamma_{\rm rel}=(\Gamma+1)/2$, 
a value which is the limit of $\Gamma_{\rm eff}$
in the ultrarelativistic domain, or, for a constant 
$\Gamma_{\rm nonrel} = \Gamma $ which is the limit of 
$\Gamma_{\rm eff}$ in the classical regime (see Fig. \ref{Fig9}). 
The consistent relativistic 
equation of state gives a wind which is always more efficient than the 
classical one corresponding to the constant value $\Gamma$, but less
efficient than the ultrarelativistic one for the value $(\Gamma +1)/2$.  
Using the value $\Gamma$ will always underestimate the asymptotic speed, but 
this difference becomes smaller as we approach the adiabatic value, as 
illustrated in Fig. \ref{Fig9}.
Conversely, it is appropriate to use  the value $(\Gamma +1)/2$ 
in the vicinity of the black hole where the temperatures are
ultrarelativistic. However it will always overestimate the asymptotic Lorentz 
factor, especially in the adiabatic case.

 As it can be seen in Fig. \ref{Fig9}b, in some cases 
the solutions can be slightly decelerated in the 
super-sonic region, because the thermal energy is not sufficient
to overcome gravity in these distances.

In the ultrarelativistic limit (large $\mu$ and $\rho$)
we can obtain analytically that our model is more efficient than the 
classical one, by comparing the
acceleration  at the sonic point for the two models.
The condition is 
\begin{equation}
\left. \frac{{d} \beta^2}{{d} R}\right|_{C} < \left. \frac{{d} \beta^2}
{{d} R}\right|_{R}
\,,
\end{equation}
 where the subscript $C$ refers to the classical polytrope,
and $R$ to the relativistic one. This is fulfilled if 
\begin{equation}
2 -\frac{1-\left.\beta_{\rm s}^2\right|_{C}}
{\left.\beta_{\rm s}^2\right|_{C}} \frac{1}{2}   
\frac{\mu}{1-\mu}>2 
-\frac{1-\left.\beta_{\rm s}^2\right|_{R}}{\left.\beta_{\rm s}^2\right|_{R}} 
\frac{1}{2}\frac{\mu}{1-\mu}
\,,
\end{equation}
which is equivalent to  
\begin{equation}
\left.\beta_{\rm s}^2\right|_{C}
< \left.\beta_{\rm s}^2\right|_{R}
\,.
\end{equation}
This condition is satisfied if the  sonic surface
 
is close enough to  the 
compact object. In the limit of large $\mu$, comparing Eqs. 
(\ref{betasonicclassic}) and (\ref{betasrho}), we conclude that
the relativistic polytrope is more efficiently accelerating the flow than the 
classical one if $\Gamma> 1$.

At this point of the comparison, 
we should stress that a constant polytropic index, as it has been 
used in many models of relativistic thermal wind, is inconsistent 
with having both ultrarelativistic and classical temperatures in the flow. 
With the usual equation of state the thermodynamic  regime is either 
ultrarelativistic or classical. In reality the outflow escapes from a very
hot corona but cools down further out so there must be a smooth transition 
from one regime to the other. 

\section{Application}

In this section we apply our model using typical values for jets from compacts
objects. We consider two cases, a supermassive black hole with  
$M=10^9\, M_{\odot}$, typical of a quasar (see \cite{Wang03}) and  
a stellar size black hole with $M= 10 \, M_{\odot}$, typical of a 
microquasar (see \cite{Mirabel98}).

\begin{figure}[h]
{\includegraphics[width=9cm,height = 8cm]{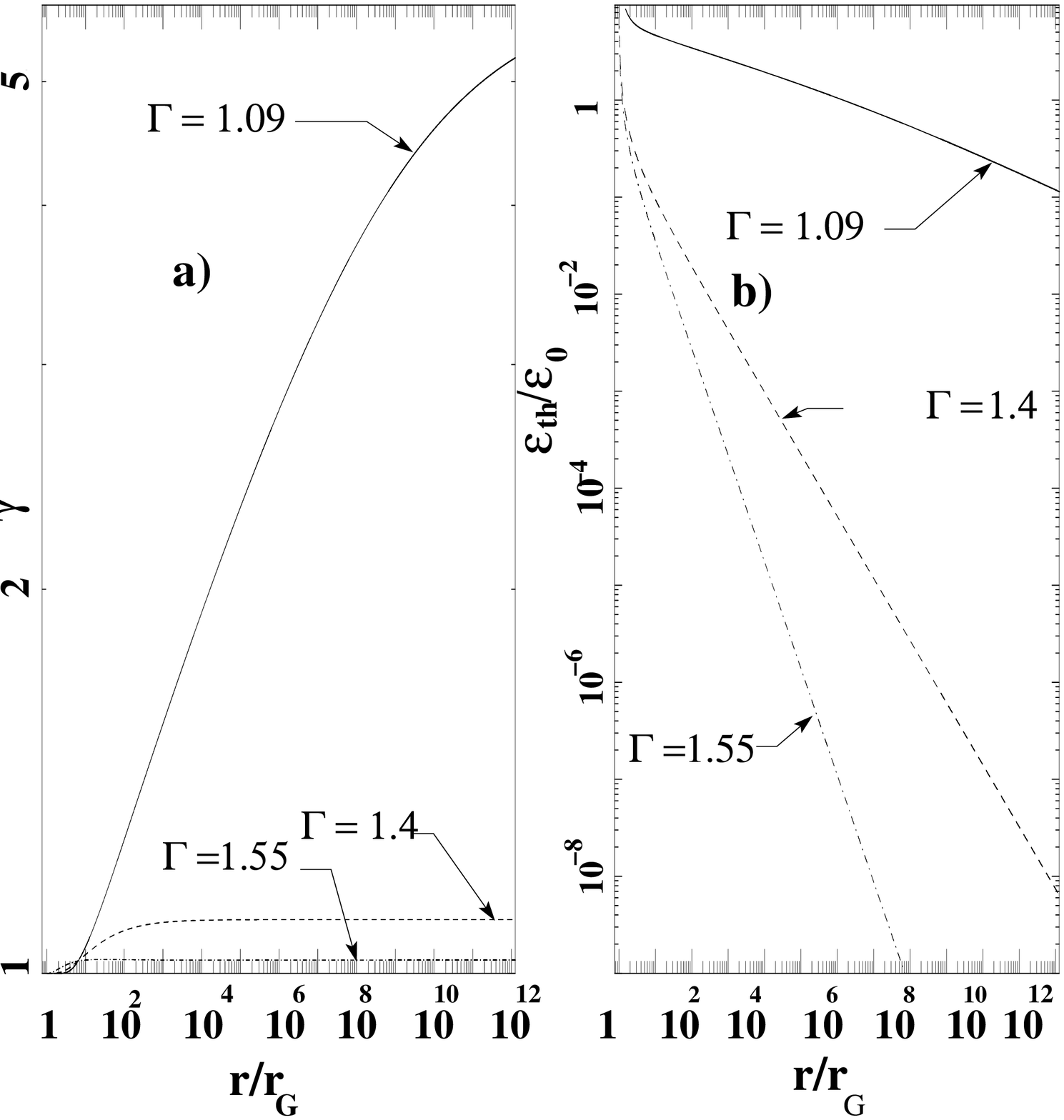}}
\caption{
Plot of the Lorentz factor  in (a) and the thermal energy density
in (b) versus radius for  $\mu = 0.155$ and three 
different values of the polytropic parameter $\Gamma$. 
}
\label{Fig10}
\end{figure}

We choose the sonic surface to be close to the compact 
object, $r_{\star} = 6.45 \,r_{\rm G}$, such that  $\mu = 0.155$. This choice  
of $\mu$ gives a sufficiently high  asymptotic Lorentz factor of 
$\gamma \geq 5$, typical of those objects for sufficiently small
values of $\Gamma$ (Fig. \ref{Fig10}).  For the solution with $\Gamma=1.09$,
the initial temperature is $T\approx 10^{12}$K in the case of electron-proton 
gas and $T\approx \,10^{9}$K in the case of electron-positron pairs.
 Note however, that for low values of $\Gamma$ such as $\Gamma=1.09$  the 
temperature remains unrealistically high at large distances. Hence, a more 
physical approach would be to avoid taking these solutions corresponding to $\Gamma \ga 1$ 
all the way to large distances, but instead match them with solutions with  
a faster decrease of the temperature ($\Gamma \approx 4/3$).

In both cases the  solutions differ  only in the density and mass loss rate. 
Both solutions have the same Lorentz factor profile and the same energy
 per particle in units of mass energy as this depends only on $\mu$ and not
on the mass of the central object or on the total mass loss rate. The result is
displayed on Fig. \ref{Fig10}.

For quasi-isothermal winds, $\Gamma\sim 1-1.2$, the Lorentz factor 
$\gamma$ can exceed a value of $5$ if 
the thermal energy distribution in the vicinity of the central object is 
roughly 10 times more than the mass energy.
The thermal content is spread along the
flow and not peaked close to the compact object such that it can accelerate
efficiently the flow with a lower maximum in temperature.
Such a quasi-isothermal outflow can occur if there is a 
redistribution of energy by the highly
radiative initial field which cools the upstream flow and reheat the
down part in the subsonic part where the medium is still optically
thick (Das 2000; \cite{Tarafdar88}). 
The dissipation of disorganized magnetic field 
could also occur to produce extended heating. Such a dissipation has 
shown to be potentially important after the  sonic surface and in the 
asymptotic region (\cite{HeinzBegelman00})

\begin{figure}[h]
{\includegraphics[width=9cm,height = 8cm]{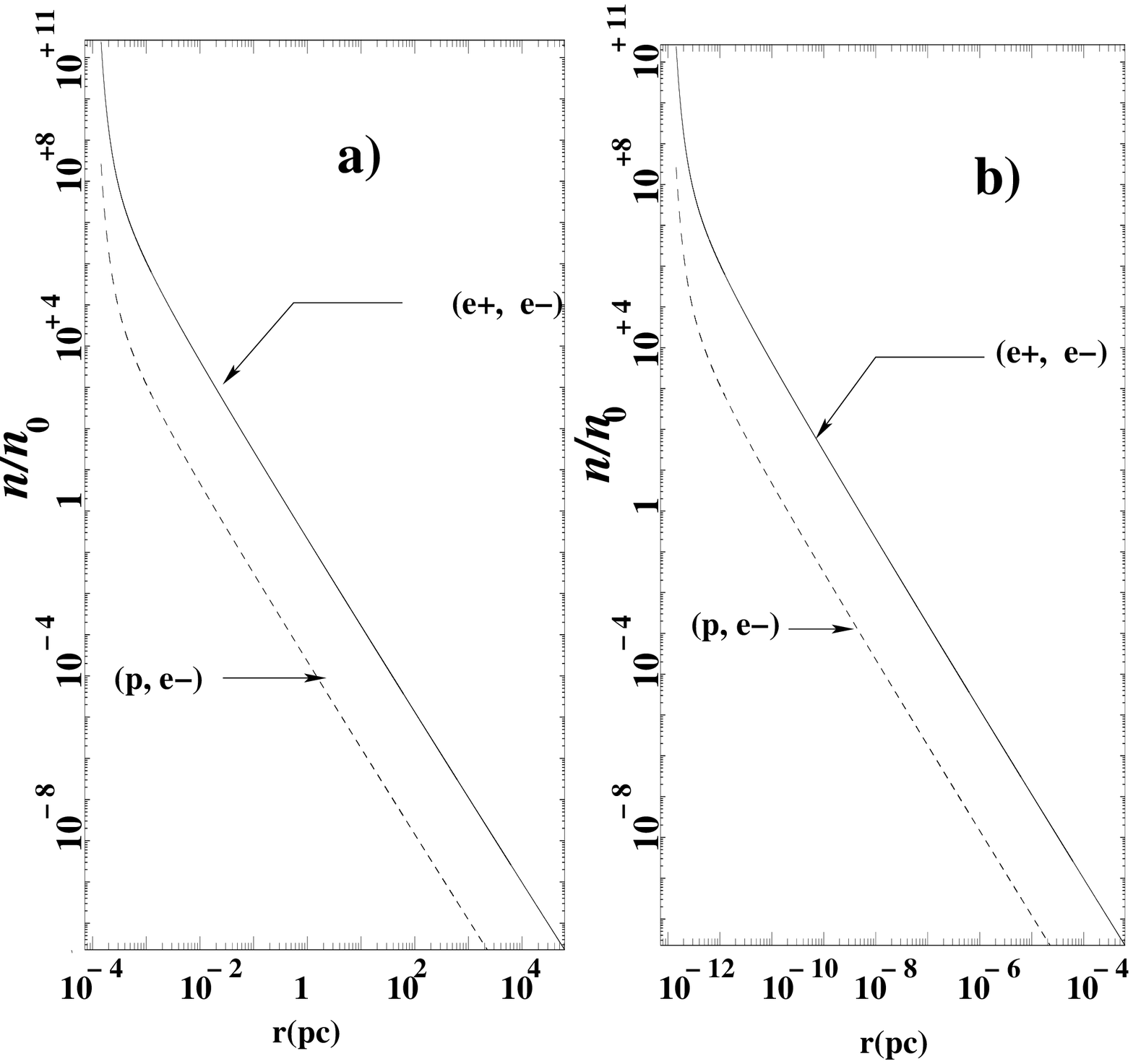}}
\caption{
Plot of density versus radius for  $\mu = 0.155$ and 
$\Gamma =1.09$
for the quasar solution in (a) and the microquasar solution in (b).}
\label{Fig11}
\end{figure}

In Fig. \ref{Fig11} we plot the number density in units of $n_0$~cm$^{-3}$ 
as a function of distance, for a free density parameter $n_0$. 
For the quasar solution (Fig. \ref{Fig11}a) the 
corresponding mass loss rate is $\dot M_{\rm quasar}
=10^{-6}n_0 \, \dot M_{\rm Eddington}=
1.41\times 10^{-6}n_0 \,M_{\odot}/$yr, while for the micro-quasar 
solution (Fig. \ref{Fig11}b) $\dot M_{\mu\rm quasar}
=10^{-14}n_0 \, \dot M_{\rm Eddington}=
1.41\times 10^{-22}n_0 \,M_{\odot}/$yr.

We note that as the mass of the central compact object changes from 
$10^9M_{\odot}$ to $ 10 M_{\odot}$  the solution is simply scaled 
down spatially, a result  consistent with the idea
that microquasars   could be considered to zeroth order as

 scaled down versions of quasars.
 A more detailed treatment however, should take into account certain differences of the 
two cases, e.g. pair production, different densities, etc., something beyond the scope of the 
present analysis.

\section{Conclusion}

We have generalized a variable polytropic index equation of state for 
the purpose of modelling relativistic flows, both in temperature and velocity 
in the
vicinity of a Schwarzschild black hole. This has enabled us to analyze  
thermally driven winds having both ultrarelativistic temperatures
at the base of the central corona ($k_{\rm B} T \ga \epsilon_{0}$)
and classical temperatures ($k_{\rm B} T \ll \epsilon_{0}$) further 
out. This equation of state is characterized 
by a polytropic index  $\Gamma$ such that pressure is related to density in
the form of Eq. (\ref{polytropePrho}).

For a given polytropic index $\Gamma$, transonic wind solutions can be found 
only within a limited range of radii of the sonic point, $r_*$, 
$$
\frac{1}{\mu_{\max}}<\frac{r_{\star}}{r_{\rm G}}<\frac{1}{\mu_{\min}}
\,. 
$$
This sonic transition should be
far enough from the Schwarzschild radius $r_{\rm G}$ such that gravity is not
too high to allow the existence of critical solutions, because for too strong
gravity the flow cannot escape, even if the temperature is infinite. 
On the other hand, the sonic transition should also be close enough 
such that the corona is not too diluted and the pressure too low to obtain 
an accelerated transonic solution.

Schwarzschild's metric tends to enhance the effects of gravity. One
major effect of strong gravity is to have a more efficient De Laval nozzle 
which allows to have accelerated winds even for polytropic indices larger 
than the typical Parker's value, i.e. $\Gamma>3/2$,
conversely to the classical  Parker's wind (\cite{Parker60}). 
This can be understood by computing the effective polytropic index 
${d}\ln P/{d}\ln \rho$. Enhanced gravity and also relativistic 
temperatures tend to lower the effective polytropic index in the low corona 
which gives a more efficient thermal driving of the wind.

However, we note that despite its widespread use in the literature,
the ordinary polytropic equation of state with a constant effective polytropic
index seems not to be really consistent with
a mixed regime of temperatures in the corona from ultra-relativistic 
to classical ones.

In order to reach very high Lorentz factors, in the adiabatic case, the 
internal energy of the plasma in the corona should exceed by a large factor 
its mass energy. 
This can be easily achieved if the total pressure is not limited to  the kinetic 
pressure of the gas but also includes extra physical processes such as MHD 
waves or radiation. This result is somehow consistent with the usual low 
velocities obtained for classical winds. Quasi-isothermal 
winds reach higher Lorentz factors with a relatively lower
-- but still larger than the mass energy -- internal energy. 

Applying our model to typical values observed in microquasars and quasars
we recover as expected that microquasar outflows may  be seen as a scaled 
down version of their bigger extragalactic counterparts. We do not claim
however that all jets are only thermally driven. 
A magnetic driving mechanism seems more efficient
indeed to accelerate the outflow on parsec scales as shown by 
Vlahakis \& K\"onigl (2004), 
in agreement with some jet observations (\cite{Sudouetal00}). 
However, we simply emphasize that thermal driving may indeed play an important 
role alongside other mechanisms, when hot coronae are observed; in such cases,
a consistent way of dealing with these relativistic temperatures is required.

Finally, this consistent generalization of the Parker polytrope for 
relativistic thermal winds could be implemented in numerical simulations, 
instead of the classical constant polytropic index equation of state to 
simulate the transition from relativistic to nonrelativistic temperatures 
along the flow. 
Nevertheless, as in any polytropic equation of state
the source of heating is not specified on physical grounds and more detailed 
physics of the coronal heating is needed.

\begin{acknowledgements}
We acknowledge financial support from the French Foreign Office
and the  Greek General Secretariat for Research and Technology
(Program Platon).  K.T. and N.V. acknowledge partial support from the
European Research and Training Networks PLATON (HPRN-CT-2000-00153)
and ENIGMA (HPRN-CT-2001-0032).
\end{acknowledgements}

\appendix

\section{Heating of polytropic flows}\label{appendix}

Conversely to the usual polytropic equation of state used in studying 
stellar interiors, the
use in classical as well as relativistic winds of a polytropic 
index different from the adiabatic one is an implicit way to mask extra 
heating in the corona as explained in
\cite{STT99}. In all cases, the fluid remains a monoatomic plasma of 
electrons and protons with a ratio of specific heats $5/3$. 

Thus, by substituting Eq. (\ref{defpolytrope}) to Eq. (\ref{monoatomic}), 
we have 
introduced implicitly that the external medium could give extra energy to the 
fluid. $e$ in Eq. (\ref{defpolytrope}) is the total internal energy of the gas.
It has two component, i.e. $e_{\rm plasma}$ the internal energy of the plasma
itself calculated from  Eq. (\ref{monoatomic})
\begin{equation}\label{monoatomic2}
e_{\rm plasma}^2 = \epsilon_{0}^2 +\kappa n^{\frac{2}{3}}
= \epsilon_{0}^2 +\kappa n^{\Gamma_{\rm ad} -1}
\,,
\end{equation}
and the heating integrated along the streamline from the source $Q$. From
a different point of view $Q$ is 
also the internal energy (defined within an arbitrary constant of course)
of the external medium (also called in thermodynamics ``universe'' as opposed
to the system) which is given to the system through heating. 

The global
system plasma $+$ external medium being isolated, it is adiabatic, so it is
usual to define $e$ instead of $e_{\rm plasma}$ in order to keep its simple
form to the energy conservation given by Eq. (\ref{Tenseur})
(e.g., \cite{BlumenthalMathews76}; \cite{MobarryLovelace86}). It is
easy to calculate the heating transfered to the plasma during its expansion,
\begin{equation}\label{heat1}
Q=e- e_{\rm plasma}
\,,
\end{equation}
which is in our case given by
\begin{equation}
Q = 
\sqrt{ \epsilon_{0}^2 + \kappa n^{\Gamma - 1}} 
- \sqrt{ \epsilon_{0}^2 + \kappa n^{(5/3) - 1}} \,.
\end{equation}

The same holds for the enthalpy. $w$ is the total enthalpy of the
gas and the external medium and if $w_{\rm plasma}$ is the real enthalpy of the
gas itself, 
\begin{equation}
w=w_{\rm plasma}+Q
\,.
\end{equation}

\end{document}